\documentclass[10pt]{article}
\usepackage{amsmath}
\usepackage{latexsym}
\usepackage{cite}
\newcommand{\be}{\begin{equation}}
\newcommand{\ee}{\end{equation}}
\newcommand{\ndt}{\noindent}
\def\bea{\begin{eqnarray}}
\def\eea{\end{eqnarray}}
\def\beas{\begin{eqnarray*}}
\def\eeas{\end{eqnarray*}}
\def\sla{\raise.15ex\hbox{$/$}\kern-.57em}

\def\Psim{{\Psi}^{-}}

\def\parm{{\partial}_{-}}

\newcommand\fr[1]{\frac{1}{#1}}

\begin{document}

\begin{titlepage}
\vskip 1cm
\centerline{\LARGE {\bf {$LC_2$ formulation of supergravity}}}

\vskip 2cm

\centerline{Sudarshan Ananth} 
\vskip .5cm
\centerline{\em  Indian Institute of Science Education and Research}
\centerline{\em Pune 411021, India}
\vskip 1.5cm

\vskip 1cm

\centerline{\bf {Abstract}}

\vskip .5cm

\noindent We formulate $(N=1, d=11)$ supergravity in components in light-cone gauge ($LC_2$) to order $\kappa$. In this formulation, we use judicious gauge choices and the associated constraint relations to express the metric, three-form and gravitino entirely in terms of the physical degrees of freedom in the theory.

\end{titlepage}

\section{Introduction}

Eleven-dimensional supergravity is an interesting theory for a number of reasons. Chief among these is that the theory is the higher dimensional progenitor for $(N=8,d=4)$ supergravity. There are indications that the $N=8$ theory is perturbatively finite making it a candidate for a finite quantum field theory of gravity - this makes understanding the parent theory important as well. In this paper, we formulate $d=11$ supergravity in components, to order $\kappa$ in light-cone gauge. This is an interesting exercise in itself: given the three very different fields in the theory, judicious gauge choices can vastly simplify the structure of the Lagrangian. We will make a number of such choices to make manifest the physical degrees of freedom and highlight the close ties between the graviton, gravitino and three-form. $(N=1,d=11)$ supergravity is ultra-violet divergent. A model like M-theory, also in eleven dimensions, presumably tames these divergences and an understanding of how may arise from a study of the ultra-violet properties of $d=11$ supergravity. Another point of interest is the role of the little group in divergence analysis. Hughes~\cite{rh} conjectured that divergence cancellations in field theories could be traced back to the space-time little group. Curtright~\cite{tc} made this proposal more concrete by considering loop integrals arising from theories in higher dimensions
\bea
\label{tc}
\Pi_{mn}(q)\propto \frac{{(-)}^s}{r}{\biggl (}I^{(2)}-\frac{I^{(0)}}{D-1}{\biggr )}(q_mq_n-q^2g_{mn})f(q^2)\ .
\eea
$f(q^2)$ represents a generic one-loop integral, $D$ the dimension of space-time and ${(-)}^s=+1$ for bosons and $-1$ for fermions. $I^{(0)}$ and $I^{(2)}$ represent Dynkin indices corresponding to $O(D-2)$ thus emphasizing the central role played by the space-time little group in determining ultra-violet behavior. The structure of $SO(9)$, for example, offers insights into the divergent nature of the $(N=1,d=11)\leftrightarrow(N=8,d=4)$ system~\cite{PMR1}. A systematic derivation of (\ref {tc}), in the context of Yang-Mills, was undertaken in~\cite{ABR0}. Extending this analysis to supergravity requires an $LC_2$ formulation in components\footnote{As opposed to a superspace or manifestly supersymmetric formulation~\cite{ABR1}.} and this motivates the present paper. A light-cone formulation where the unphysical degrees of freedom are eliminated is referred to as $LC_2$ as opposed to $LC_4$~\cite{lc4}. 

\section{Eleven-dimensional supergravity}
\label{sugras}

The bosonic field content of eleven-dimensional supergravity consists of the elfbein, ${{e_\mu}^a}$ and a completely antisymmetric 3-form potential $A_{\mu\nu\rho}$ with field strength ${F_{\mu\,\nu\,\rho\,\sigma}}={\partial_{\,[\mu}}\,{A_{\nu\,\rho\,\sigma\,]\,}}$. In terms of the $SO(9)$ little group in eleven dimensions, these correspond to a total of 128 bosonic states. The fermionic content consists of a single Majorana field, ${\Psi_\mu}$ which has 128 fermionic states. The $N=1$ supergravity action in eleven dimensions is~\cite{julia}
\bea
\label{the}
S=\int\,d^{11}x\;e\,\{{\it {L_1}}+{\it {L_2}}+{\it {L_3}}+{\it {L_4}}\,\}\ ,
\eea
where $e$ is the elfbein determinant and the individual Lagrangians are~\cite{sugra}
\bea
\begin{split}
\label{sugra}
{\it {L_1}}=&-\,\fr{2}\,e\,R(\,e\,,\,\omega\,)\ , \\\\
{\it {L_2}}=&-\fr{48}\,e\,F_{\mu\nu\rho\sigma}F^{\mu\nu\rho\sigma}+\frac{2\kappa}{(12)^4}\,\epsilon^{\mu_1\ldots\mu_{11}}\,F_{\mu_1\ldots\mu_4}\,F_{\mu_5\ldots\mu_8}\,A_{\mu_9\mu_{10}\mu_{11}}\ , \\\\
{\it {L_3}}=&-\,\frac{i}{2}\,e\,{\overline\Psi}_\mu\,\gamma^{\mu\rho\sigma}D_\rho(\frac{\omega+{\hat \omega}}{2})\Psi_\sigma\ , \\\\
{\it {L_4}}=&+\frac{\kappa}{192}\,e\,(\,{\overline\Psi}_\mu\gamma^{\mu\nu\rho\sigma\alpha\beta}\,\Psi_\beta+12\,{\overline\Psi}^\nu\gamma^{\rho\sigma}\Psi^\beta\,)\,[\,F_{\nu\rho\sigma\alpha}+{\hat F}_{\nu\rho\sigma\alpha}\,]\ .
\end{split}
\eea
We will work with both the metric and elfbein fields ${{e_\mu}^a}$, since we need to introduce fermions. We use light-cone coordinates for {\it {both}} space-time and the locally flat indices. Whenever necessary, we will circle space-time light-cone indices to differentiate them from locally flat light-cone indices. $\mu\,,\,\nu\,\ldots$ represent space-time indices, $\mu={\oplus},{\ominus},{\textcircled {i}}$ with ${\textcircled {i}}=1\,{\ldots}\,8$ while $a\,,\,b\,\ldots$ are the locally flat indices, $a=+,-,i$ where $i=1\,{\ldots}\,8$.
\vskip 0.1cm
\ndt Working in the 1.5 order formalism~\cite{deser}, we formulate the $d=11$ theory in light-cone gauge to order $\kappa$. The spin-connection, determined by the variation $\frac{\delta}{\delta\omega}S=0$ is
\bea
\begin{split}
\omega_{\nu ab}(e)&={e_a}^\rho(\partial_\nu e_{b\rho}-\partial_\rho e_{b\nu})-{e_b}^\rho(\partial_\nu e_{a\rho}-\partial_\rho e_{a\nu}) \\
&+{e_a}^\rho{e_b}^\sigma(\partial_\sigma\,e_{c\rho}-\partial_\rho e_{c\sigma}){e_\nu}^c \\
&+\frac{\kappa^2}{4}(\overline\Psi_\nu\gamma_a\Psi_b-\overline\Psi_\nu\gamma_b\Psi_a+\overline\Psi_a\gamma_\nu\Psi_b)-\frac{\kappa^2}{8}\overline\Psi^\alpha\gamma_{\alpha\nu ab\beta}\Psi^\beta\ .
\end{split}
\eea
The curvature is defined as
\bea
R_{\mu\,\nu\,a\,b}={\partial_\mu}\;{\omega_{\nu\,a\,b}}-{\partial_\nu}\;{\omega_{\mu\,a\,b}}+{\omega_{\mu\,a\,c}}\;{{{\omega_\nu}^c}_b}-{\omega_{\nu\,a\,c}}\;{{{\omega_\mu}^c}_b}\ .
\eea
Space-time $\gamma$ matrices are written in terms of locally flat coordinates as
\bea
{\gamma^\mu}={{e^\mu}_a}\,{\gamma^a}\ ,
\eea
and these flat gamma matrices satisfy
\beas
\{\gamma^a,\gamma^b\}=-2\,\eta^{ab}\ .
\eeas
$\eta^{ab}$ is flat with signature $(-1,+1,\ldots+1)$ and $\gamma^{\mu_1\ldots\mu_n}$ is the completely antisymmetric product of $n$ $\gamma$ matrices.
\bea
\label{covd}
D_\nu=\partial_\nu+\fr{8}[\gamma^a,\gamma^b]\omega_{\nu ab}\ ,
\eea
is the covariant derivative.

\section{$d=11$ supergravity in $LC_2$ to order $\kappa$}
\vskip 0.1cm
Pure gravity has been previously formulated in light-cone gauge in many mildly differing forms~\cite{scherk,kaku,bengt,schwarz,arag,ananth} while gravitino interactions have been analyzed in light-front variables in~\cite{singh}. The three fields in the $d=11$ theory are the metric, the three-form and the Rarita-Schwinger field. The metric has $44$ components, the three-form $84$ and the gravitino $128$ degrees of freedom. With the spacetime metric $(-,+,\ldots,+)$ we define
\be
x^\pm\,=\,\fr{\sqrt 2}\,(x^0\,\pm\,x^{10})\ , \quad \partial_\pm\,=\,\fr{\sqrt 2}\,(\partial_0\,\pm\,\partial_{10})\ .
\ee
$x^+$ plays the role of time and $-i\,\partial_+$ the Hamiltonian. $\parm=-\partial^+$ is now a spatial derivative and its inverse, $\frac{1}{\parm}$, is defined using the prescription in~\cite{SM}. We will now systematically formulate each of the four terms in (\ref {the}) in light-cone gauge.

\subsection{${\it L}_1$: gravity}
\label{grav}

This subsection is not as detailed as the following two because light-cone gravity has been treated in detail before. We therefore keep this subsection short highlighting only important results and refer the reader to~\cite{bengt,schwarz} for additional details. In this subsection alone, we circle the space-time light-cone indices to differentiate them from the locally flat light-cone indices.
\vskip 0.1cm
\ndt We parametrize the elfbeins as follows
\bea
{{e_\oplus}^+}={e^{{1\over 2}\,{\phi}}}\ , \qquad
{{e_\ominus}^-}={e^{{1\over 2}\,{\phi}}}\ .
\eea
The symmetric object $g_{ij}$ (transverse metric) is parametrized as
\bea
{g_{{\textcircled {i}}{\textcircled {j}}}}={e^\xi}\,{\beta_{ij}}\ ,
\eea
where $\xi$ is a real field and ${\beta_{ij}}$ is a real symmetric unimodular matrix that satisfies
\bea
{\beta_{ij}}\,{\beta_{jk}}\,=&{\delta_{i\,k}}\ .
\eea
We choose light-cone gauge by setting
\bea
\begin{split}
&{{e_{\ominus}}^+}=0\ , \nonumber \\
&{{e_{\ominus}}^k}=0\ , \nonumber \\
&{{e_{\textcircled {i}}}^+}=0\ ,
\end{split}
\eea
and $\phi={1\over 2}\,{\xi}$. We also choose~\cite{bengt}
\bea
\label{gravint}
{\beta_{ij}}={(\;e^{\,\kappa\,h}\;)}_{ij}\ ,
\eea
where $h_{ij}$ is a symmetric trace-free matrix. We expand $\beta$ as
\bea
\beta_{ij}={\delta_{ij}}+{\kappa}\,{h_{ij}}+{\mathcal O}\,({\kappa^2})\ .
\eea
To order $\kappa$, the relevant tensor contributions from (\ref {sugra}) read
\bea
T_{\mu\nu}(A)=-\fr{12}(F_{\mu\alpha\beta\gamma}F_\nu^{\alpha\beta\gamma}-\fr{12}g_{\mu\nu}F^2)\ ,
\eea
and
\bea
\!T_{\mu}^a(\Psi)\!=\!-\frac{i}{2}ee_\mu^a{\overline\Psi}_\sigma\gamma^{\sigma\rho\eta}\partial_\rho\Psi_\eta\!+\!\frac{i}{2}e{\overline\Psi}_\mu\gamma^{a\rho\eta}\partial_\rho\Psi_\eta\!+\!\frac{i}{2}e{\overline\Psi}_\sigma\gamma^{a\sigma\rho}(\partial_\rho\Psi_\mu-\partial_\mu\Psi_\rho)\ .
\eea
There are many cancelations because most contributions to order $\kappa$ occur through the transverse metric $g_{ij}=\delta_{ij}+\kappa h_{ij}$ and its inverse $g^{ij}=\delta_{ij}-\kappa h_{ij}$ which differ in sign. From the various components of $R_\mu^a-\fr{2}e_\mu^aR=T_\mu^a(A,\Psi)$ we infer that
\bea
{\xi}\,\sim\,0+{\it {O}}(\kappa^2)\ ,\qquad e\,\sim\,1+{\it {O}}(\kappa^2)\ ,
\eea
\bea
\begin{split}
{{e_{\textcircled {j}}}^-}=&-{\kappa}\,{{\partial_m}\over {\parm}}{h_{jm}}+{\mathcal O}\,({\kappa^2})\ ,
\end{split}
\eea
and
\bea
\begin{split}
{{e_{\oplus}}^-}=&-{{\kappa}\over 2}\,{{\partial_i}{\partial_m}\over {\parm^2}}{h_{im}}+{\mathcal O}\,({\kappa^2})\ .
\end{split}
\eea
The gravity Lagrangian
\bea
{\it L}_1\,\propto\,e\,e^{\rho b}e^{\sigma c}R_{\rho\sigma cb}\ ,
\eea
is now expressible entirely in terms of the physical variables and to order $\kappa$ reads
\bea
\label{gravfin}
\begin{split}
{\it L}_1=&+\fr{4}\,h_{ij}\,{\Box}\,h_{ij}+\,\frac{\kappa}{4}\,h_{ij}\,\frac{\partial_i\partial_j}{\parm^2}\,(\parm h_{mn}\,\parm h_{mn}) \\
&+\frac{\kappa}{2}\,h_{mk}(\partial_i\parm h_{mk})\frac{\partial_l}{\parm}h_{il}+\kappa\,h_{mk}(\partial_k\parm h_{im})\partial_lh_{il} \\
&+\frac{\kappa}{2}h_{mk}(\parm h_{mk})\partial_l\partial_ih_{il}+\kappa\,(\parm h_{im})(\partial_kh_{mk})\partial_lh_{il} \\
&+\frac{\kappa}{4}\,h_{ij}(\partial_ih_{kl})\partial_jh_{kl}-\frac{\kappa}{2}h_{mk}(\partial_lh_{il})\partial_kh_{im} \\
&-\frac{\kappa}{2}h_{jl}(\partial_kh_{ij})\partial_lh_{ik}+18\kappa(\frac{\partial_i\partial_p}{\parm}A_{pkl})(\parm A_{jkl})h_{ij} \\
&-6\kappa(\partial_p A_{pik})(\partial_q A_{qjk})h_{ij}+\frac{\kappa}{12}F_{iklm}F_{jklm}h_{ij} \\
&+\frac{3}{\sqrt 2}{i}\kappa\chi^{j\dagger}\gamma^{ik}\frac{\partial^k\partial^m}{\parm}\chi^mh_{ij}+\frac{3}{\sqrt 2}{i}\kappa\frac{\partial^m}{\parm}\chi^{j\dagger}\gamma^m\gamma^i\partial^p\chi^ph_{ij} \\
&-\frac{3}{\sqrt 2}{i}\kappa\chi^{j\dagger}\gamma^i\gamma^m\frac{\partial^m\partial^k}{\parm}\chi^kh_{ij}+\frac{3}{2\sqrt 2}{i}\kappa\frac{\partial^m}{\parm}\chi^{j\dagger}\gamma^m\gamma^{ik}\gamma^l\frac{\partial^l}{\parm}\chi^kh_{ij} \\
&+\frac{1}{2\sqrt 2}{i}\kappa\frac{\partial^m}{\parm}\chi^{j\dagger}\gamma^m\gamma^{ipk}\partial^p\chi^kh_{ij}+\frac{1}{2\sqrt 2}{i}\kappa\chi^{j\dagger}\gamma^{ipk}\gamma^m\frac{\partial^p\partial^m}{\parm}\chi^kh_{ij} \\
&-\frac{3}{\sqrt 2}{i}\kappa\frac{\partial^m}{\parm}\chi^{m\dagger}\gamma^j\gamma^l\partial^l\chi^ih_{ij}+\frac{3}{\sqrt 2}{i}\kappa\frac{\partial^m\partial^k}{\parm^2}\chi^{k\dagger}\gamma^m\gamma^j\parm\chi^ih_{ij}
\end{split}
\eea
\beas
\begin{split}
&+\frac{3}{\sqrt 2}{i}\kappa\frac{\partial^m}{\parm}\chi^{m\dagger}\gamma^{jk}\partial^k\chi^ih_{ij}+\frac{3}{2\sqrt 2}{i}\kappa\frac{\partial^m}{\parm}\chi^{k\dagger}\gamma^m\gamma^{jk}\gamma^l\partial^l\chi^ih_{ij} \\
&-\fr{2\sqrt 2}{i}\kappa\frac{\partial^m}{\parm}\chi^{k\dagger}\gamma^m\gamma^{jkl}\partial^l\chi^ih_{ij}+\fr{2\sqrt 2}{i}\kappa\chi^{k\dagger}\gamma^{jkl}\gamma^m\frac{\partial^m\partial^l}{\parm}\chi^ih_{ij} \\
&-\frac{3}{\sqrt 2}{i}\kappa\frac{\partial^m}{\parm}\chi^{m\dagger}\gamma^{jk}\partial^i\chi^kh_{ij}+\fr{2\sqrt 2}{i}\kappa\chi^{k\dagger}\gamma^{jkl}\gamma^m\frac{\partial^i\partial^m}{\parm}\chi^lh_{ij} \\
&+\frac{3}{\sqrt 2}{i}\kappa\chi^{k\dagger}\gamma^{jk}\frac{\partial^i\partial^m}{\parm}\chi^mh_{ij}+\fr{2\sqrt 2}{i}\kappa\frac{\partial^m}{\parm}\chi^{k\dagger}\gamma^m\gamma^{jkl}\partial^i\chi^lh_{ij}\ .
\end{split}
\eeas
It is important to note that (\ref {gravfin}) makes significant use of results (for the three-form and gravitino) derived in the following two subsections. Also, terms that explicitly involve $\partial_+$ have been eliminated by means of a suitable field redefinition - this procedure is detailed in~\cite{bengt}.

\vskip 0.5cm

\subsection{${\it L}_2$: three-form}
The 3-form part of the supergravity Lagrangian is
\bea
\begin{split}
\label{ffaction}
{\it {L_2+L_4}}=&-\,\fr{48}\,e\,F_{\mu\nu\rho\sigma}\,F^{\mu\nu\rho\sigma}+\frac{2\kappa}{{(12)}^4}\,\epsilon^{\mu_1\ldots\mu_{11}}F_{\mu_1\ldots\mu_4}\,F_{\mu_5\ldots\mu_8}\,A_{\mu_9\mu_{10}\mu_{11}} \\
&+\frac{\kappa}{96}\,e\,(\,{\overline\Psi}_\mu\,\gamma^{\mu\nu\rho\sigma\alpha\beta}\Psi_\beta+12\,{\overline\Psi}^\nu\,\gamma^{\rho\sigma}\Psi^\beta\,)F_{\nu\rho\sigma\alpha}\ ,
\end{split}
\eea
and yields the following equations of motion
\bea
\begin{split}
\partial_\mu F^{\mu\nu\rho\sigma}=&-\frac{\kappa}{576}\epsilon^{\mu_1\ldots\mu_8\nu\rho\sigma}F_{\mu_1\ldots\mu_4}F_{\mu_5\ldots\mu_8} \\
&-\frac{\kappa}{96}\partial_\alpha (\,\overline{\Psi_\mu}\gamma^{\mu\alpha\nu\rho\sigma\beta}\Psi_\beta\,)-\frac{\kappa}{8}\partial_\alpha(\,\overline {\Psi^\alpha}\gamma^{\nu\rho}\Psi^\sigma\,)\ .
\end{split}
\eea
We choose light-cone gauge by setting
\bea
\label{lc3f}
A_{-ij}=-A^{+ij}=0 \qquad A_{-+k}=A^{+-k}=0\ .
\eea
The $+kl$ component of the equations of motion determines
\bea
\label{akl1}
A^{-kl}(\kappa^0)=-\frac{\partial_i}{\parm}A^{ikl}
\eea
and
\bea
\begin{split}
\label{akl2}
&6\,\parm^2A^{-kl}(\kappa)=-\frac{\kappa}{72}\epsilon^{+-i_2\cdots i_8kl}\parm A_{[i_2i_3i_4]}\partial_{[i_5}A_{i_6i_7i_8]}-\fr{2\sqrt 2}\kappa\,\partial_p(\chi^{p\dagger}\gamma^k\chi^l) \\
&-\frac{1}{8\sqrt 2}\kappa\,\partial_p(\chi^{j\dagger}\gamma^{jpklm}\chi^m)+\frac{5}{16\sqrt 2}\kappa\,\parm(\frac{\partial^n}{\parm}\chi^{j\dagger}\gamma^n\gamma^{jklm}\chi^m) \\
&-\frac{5}{16\sqrt 2}\kappa\,\parm(\chi^{j\dagger}\gamma^{jklm}\gamma^p\frac{\partial^p}{\parm}\chi^m)+\fr{2\sqrt 2}\kappa\,\parm(\frac{\partial^p}{\parm}\chi^{p\dagger}\gamma^k\chi^l) \\
&-\frac{\kappa}{576}\epsilon^{+i_1\cdots i_8kl}\partial_{[i_1}A_{i_2i_3i_4]}\partial_{[i_5}A_{i_6i_7i_8]}\ .
\end{split}
\eea
Note that this result uses some additional information (about the gravitino) derived in the next subsection. The $LC_2$ 3-form Lagrangian is
\bea
\label{a3f}
\begin{split}
{\it {L_2}}=&-18A_{ijk}{\Box}A_{ijk}+\frac{2\kappa}{12^2}\epsilon^{+-ijkmnpqrs}{\biggl \{}
-8\,(\partial_+A_{ijk})(\partial_-A_{mnp})A_{qrs} \\
&+24\,(\frac{\partial_i\partial_q}{\partial_-}A_{qjk})(\partial_-A_{mnp})A_{qrs}-24\,(\partial_qA_{qij})(\partial_kA_{mnp})A_{qrs} \\
&+24\,(\partial_-A_{ijk})(\partial_mA_{npq})\frac{\partial_q}{\partial_-}A_{qrs}{\biggr \}}
+\frac{9}{2}\,(\partial_-A_{ikl})\partial_i[\,A^{-kl}(\kappa)\,]\ ,
\end{split}
\eea
where $A^{-kl}(\kappa)$ is given by~(\ref{akl2}). It is important to point out that ${\it L_4}$ is not included above and will be dealt with in subsection (\ref {gravi}). As in gravity, the first term in (\ref {a3f}) involves an explicit $\partial_+$ and this is easily removed\footnote{Technically, the $\partial_+$ reappears in terms involving higher orders of $\kappa$.} using a field redefinition analogous to (29) in~\cite{bengt}. $A_{\mu\nu\rho}$ has $165$ components. The first and second gauge choices in (\ref {lc3f}) eliminate $\frac{9\cdot 8}{2}=36$ components and $9$ components respectively while (\ref {akl1}) eliminates an additional $\frac{9\cdot 8}{2}=36$ components leaving us with $84$ ``physical" components for $A_{ijk}$.

\vskip 0.3cm

\subsection{${\it L}_3$: gravitino}

The gravitino-dependent terms in eleven-dimensional supergravity are (with spinor indices suppressed)
\bea
\begin{split}
\label{RSL}
{\it {L_3+L_4}}=&-\frac{i}{2}e{\overline\Psi}_\mu\gamma^{\mu\nu\lambda}D_\nu\Psi_\lambda \\
&+\frac{\kappa}{96}e(\,{\overline\Psi}_\mu\gamma^{\mu\nu\rho\sigma\alpha\beta}\Psi_\beta+12{\overline\Psi}^\nu\gamma^{\rho\sigma}\Psi^\alpha\,) F_{\nu\rho\sigma\alpha}\ .
\end{split}
\eea
As mentioned earlier, the determinant $e\sim 1+{\it O}(\kappa^2)$ and
\be
\gamma^{\mu\nu\lambda}={\biggl [} \gamma^\mu\gamma^\nu\gamma^\lambda\!-\!\gamma^\nu\gamma^\mu\gamma^\lambda\!+\!\gamma^\nu\gamma^\lambda\gamma^\mu\!-\!\gamma^\lambda\gamma^\nu\gamma^\mu\!+\!\gamma^\lambda\gamma^\mu\gamma^\nu\!-\!\gamma^\mu\gamma^\lambda\gamma^\nu{\biggr ]}\ .
\ee
We define
\bea
\label{up and down}
\gamma^\pm=\fr{2}(\gamma^0\pm\gamma^{10})\ ,\quad \gamma^+\gamma^-\Psi^\mu=2\Psi^{\mu(+)} \quad \gamma^-\gamma^+\Psi^\mu=2\Psi^{\mu(-)}\ ,
\eea
and go to light-cone gauge by setting
\bea
\label{lcgg}
\Psi_-=-\Psi^+=0\ .
\eea
We also make the additional gauge choice
\bea
\label{agc}
\gamma\cdot\Psi=\gamma^i\Psi^i-\gamma^+\Psim=0\ ,
\eea
implying that
\be
\label{fiscal}
\gamma^i\Psi^{i(-)}=0\ .
\ee
This allows us to define the ``physical" gravitino field
\bea
\label{sh}
\chi^i=(\delta^{il}+\fr{9}\gamma^i\gamma^l)\Psi^{l(-)}\ .
\eea
Equations of motion corresponding to (\ref {RSL}) are
\be
\label{RSEOM}
i\gamma^{\mu\nu\lambda}D_\nu\Psi_\lambda=\frac{\kappa}{96}\gamma^{\mu\nu\rho\sigma\alpha\beta}\Psi_\beta F_{\nu\rho\sigma\alpha}+\frac{\kappa}{8}\gamma^{\rho\sigma}\Psi^\alpha F^\mu_{\rho\sigma\alpha}\ ,
\ee
with $D_\nu$ defined by (\ref {covd}). The $\mu=+$ component implies that
\bea
\begin{split}
\label{curvonecomp}
\partial^+{\Psi^-}^{(-)}&=\partial^l{\Psi^l}^{(-)}+\frac{\kappa}{2}(\partial^jh_{jk}){\Psi^k}^{(-)}-\frac{\kappa}{2}\gamma^j\gamma^l(\partial^lh_{jk}){\Psi^k}^{(-)} \\
&+\frac{\kappa}{2}\gamma^j(\partial^+h_{jk})\gamma^m\frac{\partial^m}{\partial^+}{\Psi^k}^{(-)}-\frac{5}{8}\kappa\,\gamma^{jklm}\,\frac{\gamma^n\partial^n}{\partial_-}{\Psi^m}^{(-)}\parm A_{[jkl]} \\
&+\frac{\kappa}{16}\gamma^{jnklm}{\Psi^m}^{(-)}\,\partial_{[j}A_{nkl]}-\frac{\kappa}{2}\gamma^j{\Psi^m}^{(-)}\,\parm A_{[+jm]} \\
&+\frac{\kappa}{16}\gamma^{jk}\frac{\gamma^n\partial^n}{\parm}{\Psi^m}^{(-)}\,\parm A_{[jkm]}\ .
\end{split}
\eea
while the lower and upper components of the $\mu=i$ equation yield
\be
\label{psisolsl}
{\partial^+}\,{{\Psi^-}^{(+)}}={1\over 2}\,{\gamma^+}\,{\gamma^m}\,{{\partial^m}{\partial^k}\over {\partial^+}}\,{{\Psi^k}^{(-)}}+{\it O}(\kappa)\ .
\ee
and
\bea
\label{psip}
\begin{split}
\partial^+\Psi^{i(+)}=&+\fr{2}\gamma^+\gamma^l\partial^l\Psi^{i(-)} \\
&+\fr{4}\kappa\gamma^+\gamma^j\{(\partial^k h_{jk})\Psi^{i(-)}\}+\fr{2}\kappa\gamma^+\gamma^l\{(\partial^l h_{ij})\,\Psi^{j(-)}\} \\
&-\fr{2}\kappa\gamma^+\{(\partial^+ h_{ij})\gamma^m \frac{\partial^m}{\partial^+}\Psi^{j(-)}\}-\fr{2}\kappa\gamma^+\gamma^j\{(\partial^k h_{ij})\,\Psi^{k(-)}\} \\
&+\fr{2}\kappa\gamma^+\gamma^j\{\partial^+ h_{ij})\,\frac{\partial^l}{\partial^+}\Psi^{l(-)}\}-\frac{5}{8}\kappa\gamma^+\gamma^{ijkl}\Psi^{-(-)}\parm A_{[jkl]} \\
&+\fr{4}\kappa\gamma^{ijklm}\Psi^{m(+)}\parm A_{[jkl]}-\fr{192}\kappa\gamma^+\gamma^{ijklpm}\Psi^{m(-)}\partial_{[j}A_{klp]} \\
&-\frac{15}{8}\kappa\gamma^+\gamma^{ijkm}\Psi^{m(-)}\parm A_{[+jk]}+\fr{2}\kappa\gamma^j\Psi^{m(+)}\parm A_{[ijm]} \\
&-\fr{16}\kappa\gamma^+\gamma^{jk}\Psi^{m(-)}\partial_{[i}A_{jkm]}+\fr{16}\kappa\gamma^+\gamma^{jk}\Psi^{-(-)}\parm A_{[ijk]} \\
&-\fr{4}\kappa\gamma^+\Psi^{k(-)}\parm A_{[+ik]}+{\it O}(\kappa^2)\ ,
\end{split}
\eea
The next subsection will focus exclusively on ${\it {L_4}}$ which was also ignored earlier when deriving (\ref {a3f}). At present, we simply substitute the above results, to order $\kappa$, into the first line in (\ref {RSL}) to obtain

\bea
\begin{split}
{\it {L_3}}=&\frac{i}{\sqrt 2}\chi^{i\dagger}\gamma^0{\biggl [}\gamma^l\partial^l\Psi^{i(+)}+\gamma^+\partial_+\chi^i+\frac{\kappa}{4}\gamma^+\gamma^j\gamma^l(\partial^k h_{jk})\frac{\partial^l}{\parm}\chi^i \\
&+\frac{\kappa}{2}\gamma^+\gamma^l\gamma^p(\partial^l h_{ij})\frac{\partial^p}{\parm}\chi^j+{\kappa}\,\gamma^+(\partial_+h_{ij}){\chi^j}-\frac{\kappa}{2}\gamma^+\gamma^j\gamma^l(\partial^kh_{ij})\frac{\partial^l}{\parm}\chi^k \\
&-\frac{\kappa}{2}\,{\gamma^j}\,(\,{\partial_-}\,{h_{ij}}\,)\,{\gamma^+}\,{\gamma^m}\,{{\partial^m}{\partial^k}\over {\partial_-^2}}\,\chi^k\,{\biggr ]} \\
&-i\kappa\frac{\partial^n}{\parm}\chi^{i\dagger}\gamma^n{\biggl \{}\frac{15}{4}\gamma^{ijkl}\frac{\partial^m}{\parm}\chi^m\parm A_{jkl}-\frac{45}{4}\gamma^{ijkm}\chi^m\partial_qA_{qjk} \\
&+\frac{3}{4}\gamma^{ijklm}\gamma^p\frac{\partial^p}{\parm}\chi^m\parm A_{jkl}-\fr{8}\gamma^{ijpklm}\chi^m\partial_{j}A_{pkl} \\
&-\frac{3}{8}\gamma^{jk}\frac{\partial^m}{\parm}\chi^m\parm A_{ijk}-\frac{3}{2}\chi^m\partial_qA_{qim}+\frac{3}{2}\gamma^j\gamma^p\frac{\partial^p}{\parm}\chi^m\parm A_{ijm} \\
&+\frac{3}{4}\gamma^{jk}\chi^m\partial_{j}A_{ikm}-\frac{3}{8}\gamma^{jk}\chi^m\partial_{i}A_{jkm}+\frac{3}{8}\gamma^{jk}\chi^m\partial_{m}A_{jki} \\
&-\frac{15}{8}\gamma^i\gamma^{jklm}\gamma^p\frac{\partial^p}{\partial_-}\chi^m\parm A_{jkl}+\frac{3}{4}\gamma^i\gamma^{jpklm}\chi^m\partial_{j} A_{pkl} \\
&-\frac{3}{2}\gamma^i\gamma^j\chi^m\frac{\partial_q}A_{qjm}+\frac{3}{16}\gamma^i\gamma^{jk}\gamma^p\frac{\partial^p}{\partial_-}\chi^m\parm A_{jkm}{\biggr \}}\ ,
\end{split}
\eea
with $\Psi^{i(+)}$ given by (\ref {psip}).

\subsection{${\it {L_4}}$: gravitino three-form coupling}
\label{gravi}

We now turn to the final piece ${\it L}_4$ in (\ref {the}). A straightforward substitution of all $LC_2$ results derived thus far yields
\bea
\begin{split}
\fr{\sqrt 2}{\it {L_4}}=&-\frac{45}{4}\frac{\partial^m}{\parm}\chi^{i\dagger}\gamma^m\gamma^{ijkl}\chi^l\partial_qA_{qjk}+\frac{45}{4}\chi^{i\dagger}\gamma^{ijkl}\gamma^m\frac{\partial^m}{\parm}\chi^l\partial_qA_{qjk} \\
&-\frac{3}{2}\chi^{i\dagger}\gamma^{ijkpl}\chi^l\partial_+A_{jkp}-\frac{15}{4}\frac{\partial^i}{\parm}\chi^{i\dagger}\gamma^{jkpl}\gamma^m\frac{\partial^m}{\parm}\chi^l\parm A_{jkp} \\
&+\frac{3}{4}\frac{\partial^m}{\parm}\chi^{i\dagger}\gamma^m\gamma^{ijkpl}\gamma^n\frac{\partial^n}{\parm}\chi^l\parm A_{jkp}-\frac{9}{2}\chi^{i\dagger}\gamma^{ijkpl}\chi^l\frac{\partial_j\partial_q}{\partial_-}A_{qkp} \\
&+\frac{15}{4}\frac{\partial^i\partial^m}{\parm^2}\chi^{i\dagger}\gamma^m\gamma^{jkpl}\chi^l\parm A_{jkp}+\frac{3}{2}\frac{\partial^i}{\parm}\chi^{i\dagger}\gamma^{jkmpl}\chi^l\partial_jA_{kmp} \\
&-\frac{15}{4}\chi^{i\dagger}\gamma^{ijkp}\gamma^m\frac{\partial^m\partial^l}{\parm^2}\chi^l\parm A_{jkp}+\frac{15}{4}\frac{\partial^m}{\parm}\chi^{i\dagger}\gamma^m\gamma^{ijkp}\frac{\partial^l}{\parm}\chi^l\parm A_{jkp} \\ 
&-\frac{3}{2}\chi^{i\dagger}\gamma^{ijkmp}\frac{\partial^l}{\parm}\chi^l\partial_jA_{kmp}-\fr{8}\frac{\partial^n}{\parm}\chi^{i\dagger}\gamma^n\gamma^{ijkmpl}\chi^l\partial_jA_{kmp} \\
&-\fr{8}\chi^{i\dagger}\gamma^{ijkmpl}\gamma^n\frac{\partial^n}{\parm}\chi^l\partial_jA_{kmp}
\end{split}
\eea

\section{Conclusions}

Eleven-dimensional space-time houses $N=1$ supergravity, the largest supersymmetric local field theory with helicity two on reduction to four dimensions. In this paper we have formulated this theory to order $\kappa$, in light-cone gauge. We have made a number of gauge choices which helped differentiate between the unwanted degrees of freedom and the actual physically relevant variables. It is interesting to note how every component of ${\it L}$ depends on all three fields thanks to the maximal supersymmetry that closely links them. One very nice thing about the $LC_2$ procedure is the way the gauge conditions make the counting of degrees-of-freedom obvious. In particular, the $128$ fermionic degrees of freedom are captured entirely by (\ref {sh}). In momentum space, these structures collapse considerably and Feynman rules are therefore the next step. A necessary next step is the Lagrangian to order $\kappa^2$ which will then make an explicit check of (\ref {tc}) for the $(N=1,d=11)\leftrightarrow(N=8,d=4)$ system possible.
 
\vskip 1cm
 
\ndt {\bf {Acknowledgments}}\\[0.5cm]
I am grateful to Sarthak Parikh and Hidehiko Shimada for valuable discussions and Martin Cederwall for sending me a replacement copy of~\cite{bengt}. I thank Lars Brink, Stefano Kovacs, Sunil Mukhi, G Rajasekaran, R Ramachandran, Pierre Ramond and Sergei Shabanov for helpful comments. This work is supported by the Department of Science and Technology, Government of India through a Ramanujan Fellowship and by the Max Planck Society, Germany through the Max Planck Partner Group in Quantum Field Theory.

\end{document}